# Journal Maps, Interactive Overlays, and the Measurement of Interdisciplinarity on the Basis of Scopus Data (1996-2012)

Loet Leydesdorff,[a]* Félix de Moya-Anegón,[b] and Vicente P. Guerrero-Bote[c]

**Abstract**

Using Scopus data, we construct a global map of science based on aggregated journal-journal citations from 1996-2012 (*N* of journals = 20,554). This base map enables users to overlay downloads from Scopus interactively. Using a single year (e.g., 2012), results can be compared with mappings based on the Journal Citation Reports at the Web-of-Science (*N* = 10,936). The Scopus maps are more detailed at both the local and global levels because of their greater coverage, including, for example, the arts and humanities. The base maps can be interactively overlaid with journal distributions in sets downloaded from Scopus, for example, for the purpose of portfolio analysis. Rao-Stirling diversity can be used as a measure of interdisciplinarity in the sets under study. Maps at the global and the local level, however, can be very different because of the different levels of aggregation involved. Two journals, for example, can both belong to the humanities in the global map, but participate in different specialty structures locally. The base map and interactive tools are available online (with instructions) at http://www.leydesdorff.net/scopus_ovl.

[a] Amsterdam School of Communication Research (ASCoR), University of Amsterdam, Kloveniersburgwal 48, 1012 CX Amsterdam, The Netherlands; loet@leydesdorff.net ; * corresponding author.
[b] CSIC, SCImago Research Group, Centro de Ciencias Sociales y Humanas, Instituto de Políticas y Bienes Públicos, Calle Albasanz 26, Madrid 28037, Spain; felix.moya@scimago.es .
[c] Universidad de Extremadura, Information and Communication Science Department, SCImago Research Group; Pl. Ibn Marwan, Badajoz, 06071, Spain; guerrero@unex.es

## 1. Introduction

Since we explored the possibilities for mapping Scopus data in terms of aggregated journal-journal citations (Leydesdorff, de Moya-Anegón, & Guerrero-Bote, 2010), the technique for generating *interactive overlays* to global maps of science (e.g., Boyack *et al*., 2005; Rafols *et al*., 2010) has been further developed, mainly for data derived from the Web-of-Science (WoS) database of Thomson Reuters. Interactive overlay maps enable users to project a set of documents onto a base map in terms of the journal distribution in the download. Using a global map of journals, one can assess the portfolio in terms of the spread across journals and journal categories, and also measure "interdisciplinarity" in terms of the journal coverage of the set(s) under study.

With the added value of using VOSviewer for the visualization of large sets, Leydesdorff, Rafols, and Chen (2013) recently organized the combined Journal Citation Reports 2011 of the Science and Social Sciences Citation Indices (SCI and SSCI, respectively) into a single file that can be mapped and interactively overlaid with sets of documents downloaded from WoS. This routine uses the distances on the map—in addition to the distribution—for the specification of Rao-Stirling diversity as a measure of interdisciplinarity in the document sets under study (Porter *et al*., 2007; Rafols & Meyer, 2010; Rao, 1982a and b; Stirling, 2007; Zhou *et al*., 2012). Rao-Stirling diversity accounts not only for the variations in set(s), but also for the ecological distances among subsets.

Rafols, Porter, & Leydesdorff (2010) first used WoS Subject Categories for developing interactive overlays on base maps. The 220+ Subject Categories—or equivalently in the case of Scopus, the 300+ so-called Minor Subject Areas—can be mapped more clearly, but at the price of indexer effects which are not needed when one can use journal-journal citations (Rafols & Leydesdorff, 2009). The journals group naturally in the network of aggregated citation relations, and thus shape an ecology. In this study, we report on our attempt to construct a similar overlay at the journal level to the Scopus database of Elsevier.

Compared with WoS data, the Scopus database has both advantages and disadvantages. One major advantage of Scopus is the inclusion of journals in the arts and humanities into a single set, and a broader coverage of the journal literature in general. The complete set of Scopus for the period 1996-2012 covered 20,554 source journals, as against 10,936 journals in WoS for 2011 as a single year.[1] One disadvantage of using Scopus for overlays is the limit of 2,000 on downloading a collection of retrieved records, while one can download up to 100,000 records in WoS (in batches of 500). Scopus does not provide a database equivalent to JCR in WoS. We used the entire set of 1996-2012 instead because this large set can be expected to provide us with a base map as reliable as possible. (Otherwise, the citation density in more peripheral regions of the map may be insufficiently populated given the broad scope of Scopus data.)

Using Scop2WoS.exe (available at http://www.leydesdorff.net/scopus ) one can already export records retrieved from Scopus in a format similar to WoS and then use previously developed software for the journal mapping of WoS data (available at http://www.leydesdorff.net/journals11). However, the titles of journals are abbreviated differently in the two databases, and thus this route may require substantial editing of the journal titles or the standardization of abbreviations by the user. Furthermore, we were interested in the shape of the base map using Scopus data.

[1] 8,471 journals are covered by the SCI, and 3,047 by SSCI in 2012, but there is an overlap of 582 journals covered by both databases (8,471 + 3,047 – 582 = 10,936).

Since two of us have been deeply involved with the Scopus data for a number of years (Moya-Anegón *et al*, 2007; González-Pereira, Guerrero-Bote & Moya Anegón, 2010; Lancho-Barrantes, Guerrero-Bote & Moya Anegón, 2010; Guerrero-Bote & Moya Anegón, 2013), and the other author developed the overlay to WoS data, we thought it worthwhile to combine our efforts in a project to make a mapping similar to that available for WoS, but using Scopus data. How do the maps compare? Can one use both for the measurement of interdisciplinarity when defined as Rao-Stirling diversity? Do the Scopus maps add to our insight into the structure of the journal literature?

A limitation to the use of journals for the mapping and the indication of interdisciplinarity is the interdisciplinarity *within* journals. Large journals such as *PLoS ONE* may be deliberately interdisciplinary. In response to this, several teams have embarked on clustering at the level of articles (Boyack *et al.*, 2014; Waltman & Van Eck, 2012). This can lead to detailed maps, but maps at this level of detail cannot be used for interactive overlays because new publications cannot be categorized such as in the case of journals. Each position of an article is then unique. Our primary objective is to provide a user-friendly interface that allows for studying one's own datasets in relation to the structure of science provided by the journal map.

## 2. Methods and data

The design is kept as similar as possible to the previous mapping of WoS data (Leydesdorff, Rafols, and Chen, 2013); we combine parts of this routine with previous efforts to map Scopus data (2008) in terms of local maps (Leydesdorff, de Moya-Anegón, and Guerrero-Bote, 2010). However, the previous maps of Scopus were not interactive.

### *2.1 Data*

Two files were extracted from the entire Scopus database (1996-2012): one with the 20,554 journal titles and other unique journal characteristics such as the total numbers of citations, references, and self-citations; the other with 14,378,017 values for the aggregated citation relations among these journals (including a record of self-citations for each journal). This information is organized into a relational database management system that was previously developed by Leydesdorff and Cozzens (1993) for the purpose of mapping JCR data.

Among the ($20,554^2 = 422,466,916$) possible relations in this grand matrix, only 14,378,017 cells are filled. This is approximately 3.4%, whereas Leydesdorff *et al.* (in press) provide a value of 1.94% for JCR data. The difference is caused by the practice in JCR to sum long tails (of single citations) under the heading of "all others."[2] In the Scopus data, however, 6,027,429 (41.9%) of the links are single citation relations. We did not include these relations, but only the 8,350,588 (58.1%) links with more than a single citation relation. The removal of these single citation relations can be expected to have an effect on network parameters such as the largest component and the density. However, the percentage of cells with a value in the matrix is then 1.97%, and thus consistent with the value of 1.94% reported for WoS in the 2011 sample.

[2] Our routines declare this data as missing values.

The sum total of citation relations in the grand matrix is 270,115,991, or after discarding the more than six million single occurrences, 264,088,562. This number of citations is an order of magnitude larger than in WoS because of the compilation across a longer period (1996-2012); Leydesdorff *et al.* (2013) reported 35,295,459 citations in WoS for 2011. For the more precise comparison we also generated a one-year database for 2012 (by setting a filter). In this case, we obtain approximately 39 million citations among the approximately 19,000 journals. However, we will not use this set for the base map because the tenfold larger numbers of the set for 1996-2012 lead to greater robustness of the base map in the details. A disadvantage is that changes in relations among journals are not visible from this static design; one can imagine a design of evolving base maps.

| | *Scopus 1996-2012* | *Scopus 2012* |
|---|---|---|
| *Journals* | 20,554 | 18,595 citing<br>19,725 cited |
| *Citations* | 270,115,991 | 38,845,698 citing<br>39,177,055 cited |
| *Citation links*<br>*(single citation relations)* | 14,378,017<br>(6,027,429) | 6,672,033<br>(3,628,012) |

**Table 1**: Descriptive statistics of the input files.

Whereas in the case of JCR the domain of source journals is defined very strictly in terms of journals, this delineation seems looser in the case of Scopus. Dissertations from major American universities (e.g., “Cornell University, Dissertation”) are also included, as are seven *Procedia* volumes that Elsevier generates from conferences in various disciplines. We did not change this data, but accepted the delineations and definitions of the Scopus staff.

### *2.2 Visualization*

For a long time, the problem with global maps has been the cluttering of the (> $10^4$) labels on the map. As noted, this was a major reason to use higher-level aggregates such as WoS Subject Categories (Rafols *et al.*, 2010). However, this problem of too many labels on a single map has been solved by both VOSviewer (at http://www.vosviewer.com) and Gephi (at https://gephi.org/). Gephi has the advantage of combining the visualization with graph-theoretical algorithms that are used in social-network analysis (e.g., centrality measures, community-finding algorithms, etc.). Gephi visualizations can also be brought online using the Gexf format (at http://gexf.net/format/). The labels in the visualization can be set proportionally in Gephi, but reading the tiny labels may require considerable zooming of the scalable visualizations before one obtains sufficient resolution (Leydesdorff, Hammarfelt, & Saleh, 2011).

VOSviewer uses a technique of fading the less important labels in a network, but the faded labels remain available in reaction to hovering with the mouse or zooming. One can also adjust the label sizes and/or the variation among the sizes. This solves the problem of the cluttering of the labels (Van Eck & Waltman, 2010). Whereas Gephi provides options to make interactivity possible in the future at the Internet (using GEXFExplorer), currently VOSViewer in our opinion is the best choice for two reasons: (1) the problem of cluttering of the labels is solved in VOSViewer, and (2) the layout is based on stress-minimization as in multidimensional scaling (MDS) whereas Gephi uses a spring-embedder (Fruchterman & Reingold, 1991). The difference between these two approaches is specified in Leydesdorff (2014) and Leydesdorff & Rafols (2012).

The output of VOSviewer can be webstarted; the user in this case has also access to all options for embellishing the resulting maps. More recently, VOSviewer also became more integrated into the network statistics provided by Pajek v3. Actually, we use the Pajek format for the exchange, and compare the results of the clustering algorithm of VOSviewer (Waltman *et al.*, 2010) with those of using the algorithm of Blondel *et al.* (2008) for the decomposition—and hence the coloring. Among the many algorithms that are currently available for community detection (Fortunato, 2010), these two are implemented in the environments of Pajek and VOSviewer that were used for this construction. However, the colors used for the nodes in the overlay maps can be made compatible with any classification if the user so wishes.

Leydesdorff *et al.* (2013) used VOSviewer for the mapping of 10,675 journals in JCR 2011 of WoS. Using a cosine threshold of 0.2 in that study, we were able to generate a base map (both cited and citing) using 8 GB in a laptop computer (under Windows-7 with a 64-bits operating system). As could be expected, the much larger-sized database of Scopus 1996-2012 led to problems following this routine, so that we had to upscale to a larger Unix-based machine in which 24 GB was available for the processing.

The cosine normalization of this huge matrix also led to problems using SPSS (v21). However, we first explored the non-normalized affiliations (or co-occurrence) matrices, both cited and citing, since these can be generated from the asymmetrical citation matrix by using Pajek. Since the results were not convincing, we had to solve the problem of cosine-normalization (using Pajek), but this will be pursued only along the "citing side" of the matrix. As in the previous study, this side is richer than the cited one because "citing"—with references to communalities in the knowledge base—provides the variation in the current year, whereas “cited”—shared citation impact—accumulates in terms of the established structures among larger (and sometimes archival) journals. One would therefore expect the “citing” map to be more informative. As we shall see below, one risks having too much variation on the citing side and too little variation (or, in other words, too much structure) on the cited.

Fortunately, base maps have to be made only one single time; the overlays can thereafter be based on the coordinates provided by the base map. Base maps of very different quality in terms of underlying data can therefore still be compared in terms of their usefulness at interactive interfaces. We use the base maps for two purposes:

1. as the basic framework on which the journal distribution of a set downloaded from Scopus can be projected; these maps can be used for portfolio analysis or for comparisons among different sets; and
2. as a distance map for measuring Rao-Stirling diversity in the sets under study; Rao-Stirling diversity can be considered as a measure of interdisciplinarity—in this case in terms of the journal composition.

### *2.3 Interdisciplinarity*

Rao-Stirling diversity was first proposed by Rao (1982a and b) in the context of mathematical biology; Stirling (2007) suggested using it as a general measure for analyzing diversity in science, technology, and society. Rafols & Meyer (2010) elaborated this measure as one among three different operationalizations of “interdisciplinarity” (Leydesdorff & Rafols, 2011; Wagner *et al.*, 2011).

Rao-Stirling diversity is defined as follows:

$$\Delta = \sum_{ij} p_i p_j d_{ij} \tag{1}$$

where $p_i$ and $p_j$ are proportional representations of the journals *i* and *j* in the system and $d_{ij}$ is the degree of difference (disparity) between journals *i* and *j* on the map. As Stirling (2007, at p. 712) argues "the simplest way to conceive of disparities between elements is as a distance between points in disparity space." The distance $d_{ij}$ in Eq. 1 is in this study defined as the Euclidean distance on the base map $\| x_i - x_j \|$ between each two journals participating in the set as a proportion of the maximally possible distance (that is, the diagonal of the map).[3]

This distance measure is an optimization[4] and projection in two dimensions (*x* and *y*) of the multi-dimensional proximities (*cosine*) among journals. Leydesdorff, Kushnir, & Rafols (2012, in press) used (1 – *cosine*) as a distance measure for an analogous mapping of (USPTO) patents in terms of International Patent Classifications (IPC). However, the number of IPC classes at the four digit-level was only 637, whereas the number of journals under study in this case is more than 20,000. The number of distances would therefore be on the order of $10^8$. Even after setting a threshold, the size of the files and therefore the processing times would remain too large for interactive use at the internet.

[3] The diagonal of the map is determined by using the largest and smallest values in the *x*- and *y*-dimensions, respectively, for drawing a square.

[4] In the case of MDS, a stress value of the projection is minimized. VOSviewer can be considered as a variant of MDS used for the visualization (Leydesdorff & Rafols, 2012; Van Eck & Waltman, 2010).

### *2.4 The research process*

The project was developed in stages:

a. First, we organized the data so that local maps could be produced using techniques similar to those used for existing maps of JCR data (Section 3.1); for reasons of comparison, we did this both for 2012 as a single year and for the aggregated set 1996-2012;
b. Because of the large size of the grand matrix (1996-2012), we first developed a non-normalized citation matrix (in both the cited and citing directions) and visualized these (Section 3.2);
c. Since the results of the non-normalized matrix were unsatisfactory, we cosine-normalized this matrix on the citing side (Section 3.3), and generated an overlay system for the mapping of samples—that is, retrieved document sets (Section 3.4). This system will be illustrated in comparison to results reported previously for JCR data, and using, for example, data from the humanities (Section 4);
d. A website was developed at http://www.leydesdorff.net/scopus_ovl where the user can generate overlays from any set(s) downloaded from Scopus (at http://www.scopus.com); the Rao-Stirling diversity of a set under study is stored in a file "rao.txt" on the occasion of each subsequent analysis. An instruction is provided in the Appendix.

## 3. Results

### *3.1 Local maps*

As an example of the differences between Scopus and JCR-WoS, Figure 1 shows the local maps based on the ego-networks of the *Journal of Informetrics* to the extent of 0.5% of the journal's total citation rate in Scopus 2012 and JCR 2011, respectively. The numbers of journals in these ego-networks are 21 and 15, respectively. Using Blondel *et al.*'s (2008) algorithm, three communities are distinguished in both cases with modularity $Q = 0.197$ and $Q = 0.057$, respectively. The left-side figure based on Scopus data shows a very different (and larger) set of journals than the right-side figure based on WoS data.

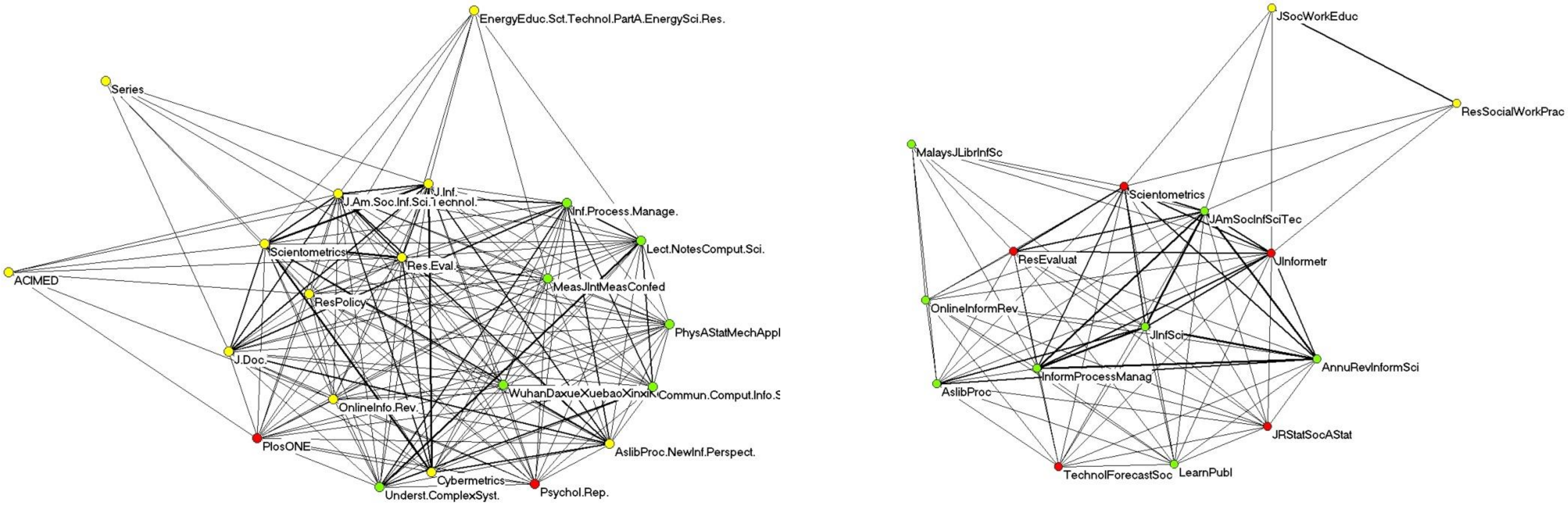

**Figure 1**: *J of Informetrics* (*J. Inf.*) cited in the relevant citation environment in Scopus 2012 (left side) and JCR-WoS 2011 (right side). Only those journals were included which contribute more than 0.5% to the total citations of the seed journal (*J. Inf.*); no cosine threshold. In the case of Scopus (left): *N* of journals = 21; *N* of Communities = 3; *Q* = 0.197; in JCR-WoS: *N* of journals = 15; *N* of communities = 3; *Q* = 0.057 (Blondel *et al.*, 2008); Kamada & Kawai (1989) used for the visualization.

As expected, the richer environment of Scopus in terms of the numbers of journals included can lead to more detailed representations of citation environments in local maps.[1] However, the domains of the databases are different, and therefore also the maps. The disturbances because of the inclusion of more marginal journals or non-journals in Scopus can be filtered out by setting a (relatively low) threshold. Both databases provide informative maps, albeit sometimes rather different in content.

### *3.2. A global map based on non-normalized citation relations* ($N$ = 20,554)

Since the cosine-normalization of the large file of aggregated journal-journal citation relations generated memory problems, we first explored the data using the non-normalized data as input to VOSviewer. The asymmetrical citation matrix was to this end transformed into a symmetrical 1-mode affiliations matrix after reading the data into Pajek. The implied multiplication with the transposed can be performed along both axes of the matrix ("cited" and "citing", or, in formal language: $A*A^T$ or $A^T *A$).

| | *Cited* | *Citing* |
|---|---|---|
| *Largest component* (N) | 19,140<br>(93.1%) | 19,604<br>(95.4%) |
| *N of clusters* (Blondel *et al.*, 2008) | 6 | 12 |
| *Modularity Q* | 0.400 | 0.591 |
| *N of clusters* (VOSviewer) | 10 | 814 |

**Table 2**: Affiliations matrices in the cited and citing dimensions ($N$ = 20,553).

[1] We did not set a threshold to the cosine values because that leads to isolates and thus may impair the possibility of comparison.

Table 2 shows that the variation in the citing dimension leads to a proliferation in the clustering when using VOSviewer; the number of clusters is then 814. The maps, however, are not so different in the cited and citing dimensions. Figure 2 provides the "cited" map which shows the disciplinary structure of the sciences in terms of the major journals, mainly in the domains of the physical and life sciences. Journal names in the domains of the social sciences and humanities are completely overshadowed (because the data were not yet normalized for size).

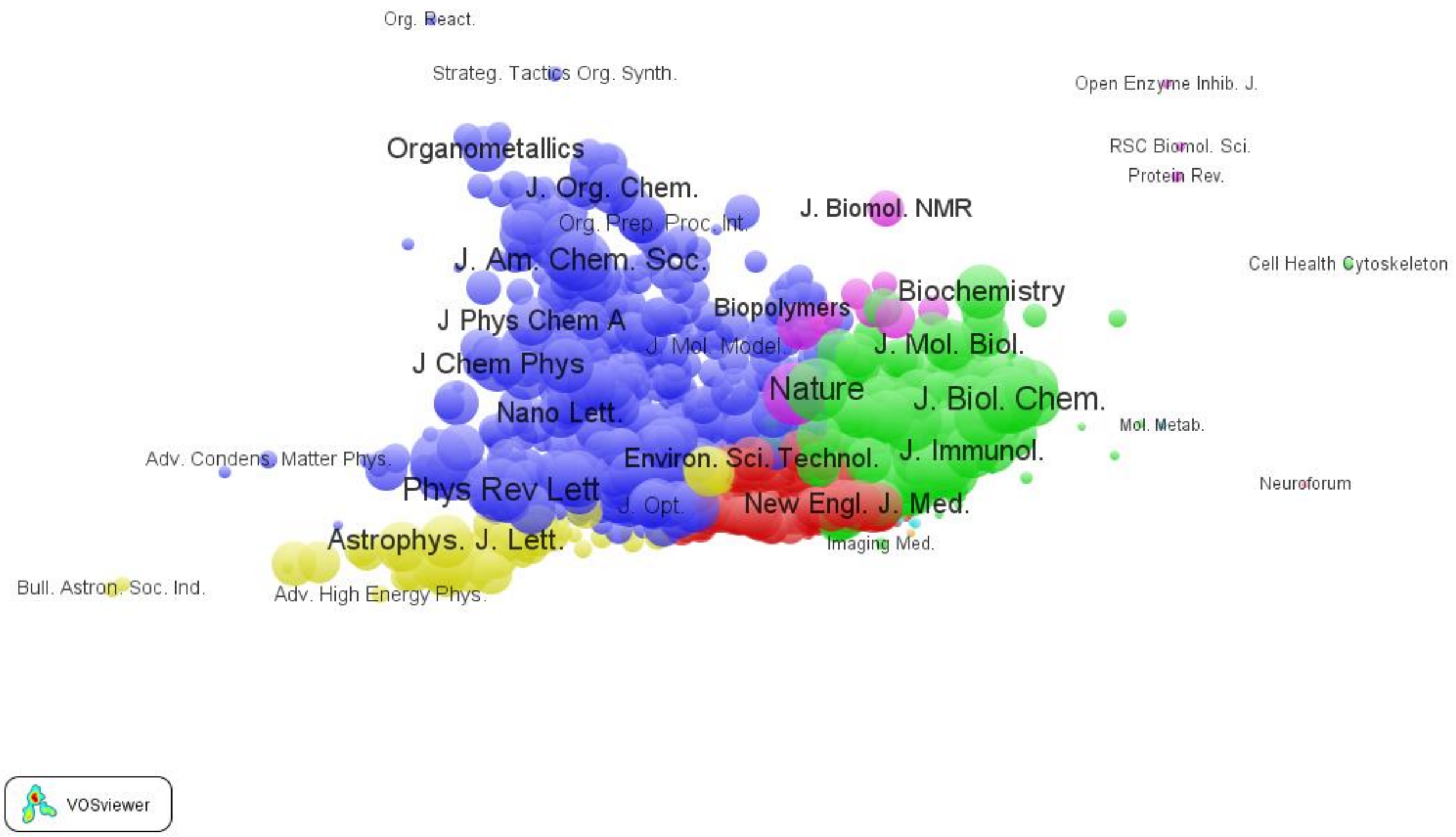

**Figure 2**: 10 clusters distinguished by VOSviewer in the non-normalized aggregated citation patterns of 19,140 cited journals. Node sizes and labels are rescaled. This map is available at http://www.vosviewer.com/vosviewer.php?map=http://www.leydesdorff.net/scopus_ovl/cited.txt&label_size=1.30&label_size_variation=0.13

This map shows five main disciplinary groups, such as chemistry, physics, bio-medicine, etc. (The five other colors indicate groups of fewer than five journals.) Without resizing of the labels and the nodes, major journals such as *Journal of Biological Chemistry, Journal of the American Chemical Society*, *Physical Review Letters*, etc., dominate the representation. This is also the case for the map based on the citing patterns, but then the clustering fails to converge sufficiently in VOSviewer (cf. Waltman *et al*., 2010): among the 814 clusters distinguished, only 76 contain five or more journals.

In summary, the map shows a dense packing of the database in terms of major journals that are cited pervasively. Cosine-normalization can be expected to improve on this visualization considerably. Furthermore, the citing dimension provides us with the activity also of minor journals because "citing" is pervasive in the database, whereas "being cited" is heavily concentrated in the major journals.

### *3.3. A global map based on cosine-normalized citation relations*

The above results made it worthwhile to solve the problem of generating cosine-normalized maps for these large files. Given the co-occurrence values used in the previous routine, one can divide the numerator of the cosine by the quadratic summations that can be added as a journal characteristic to each of the 20,554 journals. The cosine then follows as a quotient, as follows:

$$\mathrm{Cosine}(x,y) = \frac{\sum_{i=1}^{n} x_i y_i}{\sqrt{\sum_{i=1}^{n} x_i^2}\sqrt{\sum_{i=1}^{n} y_i^2}}$$

We focus the presentation on the “citing side” for the reasons mentioned above. The resulting file contains the largest component of 19,603 journal titles, or 95.4% of the 20,554 in the original file (Table 2).

Initial results using this largest component in VOSviewer showed an informative map; but three (extreme) outliers stretched the map: *Hospitals Health Networks*, *Parish Nurse Perspectives,* and *Metascience*. After removing these three nodes, the number of communities using Blondel *et al*. (2008) was 36 ($Q$ = 0.667). We entered this file ($N$ = 19,600) again into VOSviewer without the partition information so that the clustering results generated by this program were additionally obtained.

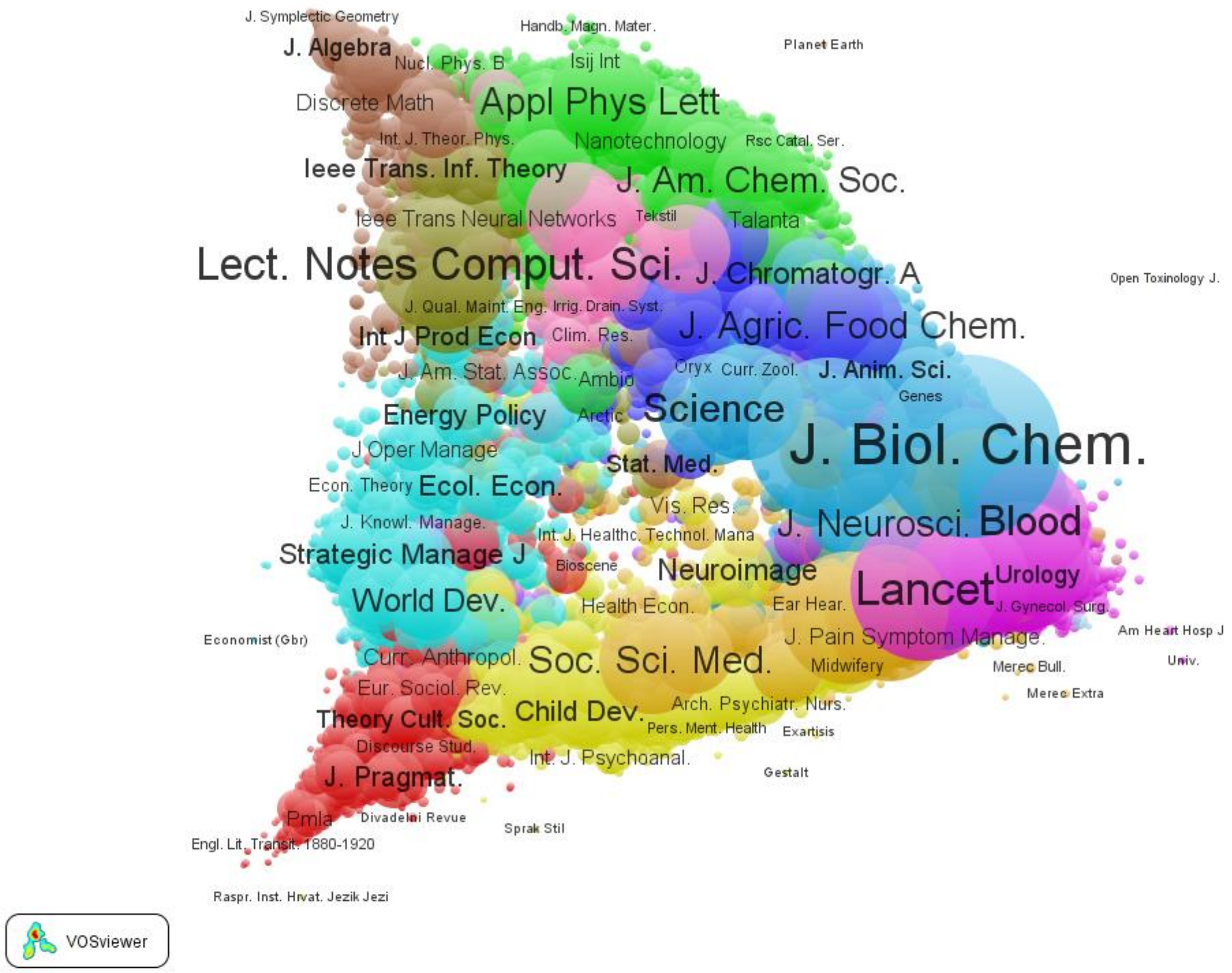

**Figure 3**: Base map of aggregated citation relations among 19,600 journals included in Scopus 2012; colors correspond to 27 communities distinguished by VOSviewer; available at http://www.vosviewer.com/vosviewer.php?map=http://www.leydesdorff.net/scopus_ovl/basemap.txt

This map is based on all cosine values, that is, *without a threshold*. VOSviewer distinguishes 27 communities, of which 16 contain more than two journals. In our opinion, the map is informative: clockwise, one can, for example, at the top left distinguish mathematics journals, followed by journals of physics and chemistry. One can further descend on the right side of the figure to agriculture, biology, bio-medicine, and eventually the clinical sciences. To the left of these, psychology, and the social and cultural sciences span across the bottom to the left side of the figure. On the left side of the triangle, economics journals and energy research color differently. Computer science is positioned at the top of the inner triangle among physics, chemistry, and the model-oriented sciences (such as economics and energy research).

Maps as two-dimensional representations of multi-dimensional data should not be over-interpreted. VOSviewer does not provide us with stress values of the projection as indicators of uncertainty (Borgatti, 1997; Kruskall, 1964; Leydesdorff & Schank, 2008). Maps are functional for the representation and the discussion, but for analytic purposes one should probably prefer algorithmic approaches that lead to tables such as factor analysis (Schiffman *et al.*, 1981).

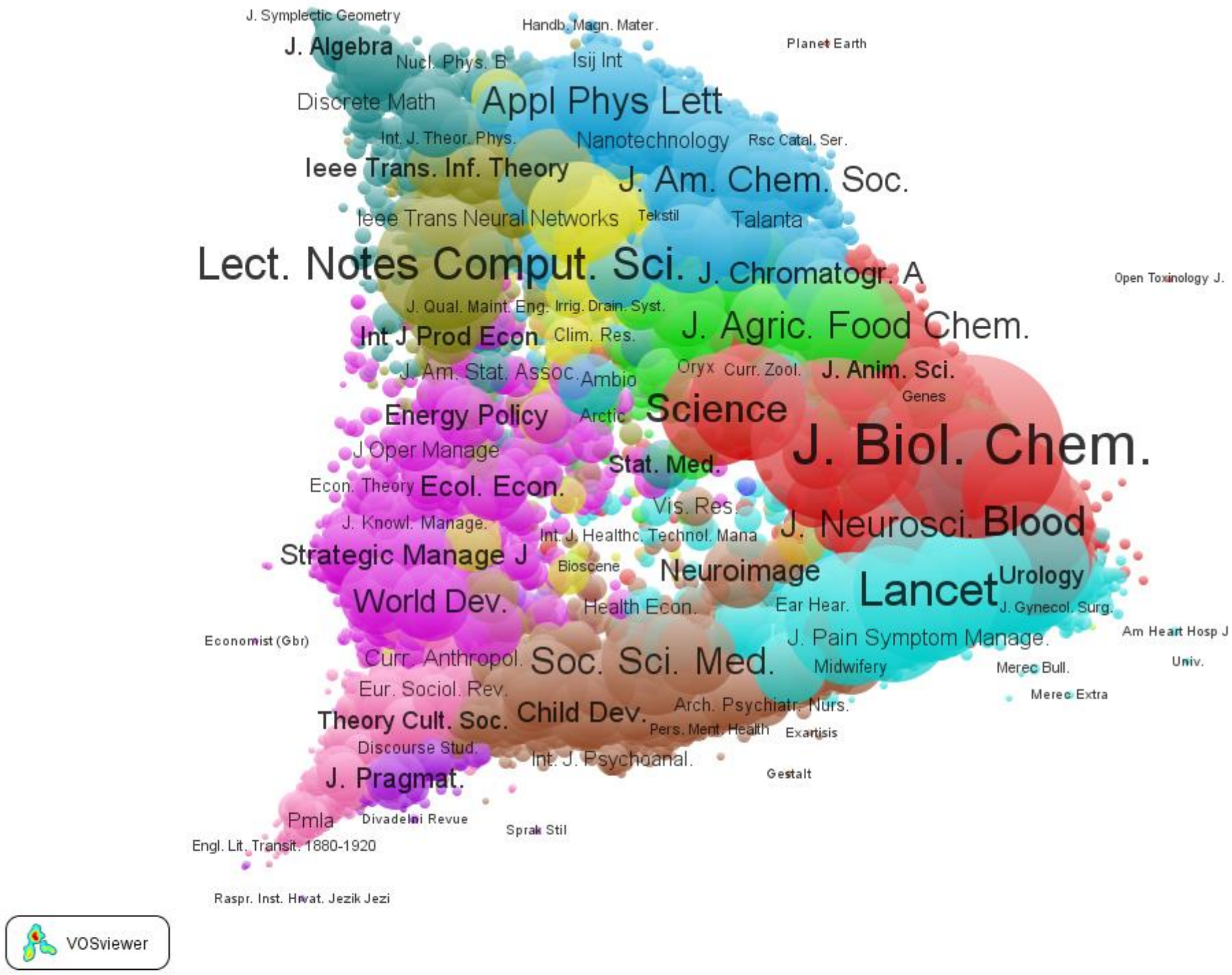

**Figure 4**: Base map of aggregated citation relations among 19,600 journals included in Scopus 2012, and colored according to the community structure ($N$ of clusters = 36; $Q$ = 0.667) generated by the algorithm of Blondel *et al.* (2008); available at http://www.vosviewer.com/vosviewer.php?map=http://www.leydesdorff.net/scopus_ovl/blondel.txt

The results of using the classification generated by Blondel *et al.*'s (2008) algorithm are shown in Figure 4. (We used this classification for the base map in WoS; cf. Leydesdorff *et al.*, 2013). Among the 36 communities distinguished, 15 contain more than one or two journals. Upon visual inspection, Figure 4 does not obviously improve on the solution of VOSviewer used in Figure 3. Both algorithms generate a relatively large number of isolates.

More detailed discussion about the specific differences between these two solutions for community finding would lead us away from the objectives of this study, that is, the generation of a base map, the construction of interactive overlays, and the measurement of interdisciplinarity. As noted, many more algorithms for this purpose are now available (Newman & Girvan, 2004; Waltman & van Eck, 2013; cf. Fortunato, 2010). Other classification systems can also be based on human indexing or mixtures between formal methods and human indexing such as the map of Science-Metrix at http://www.science-metrix.com/OntologyExplorer/ (cf. Rafols & Leydesdorff, 2009). For reasons of simplicity, we decided to take as the default the results of VOSviewer (in Figure 3) both in terms of the mapping and the clustering. However, the user can change to the classification based on Blondel *et al.* (2008) or any other community structure if so wished.[2]

### *3.4 The generation of overlay files*

The generation of overlays to the base map (in Figure 3) is possible for any download from Scopus defined by the user. One of the output formats of Scopus can be used as input to the program overlay.exe that is available online at http://www.leydesdorff.net/scopus_ovl . This routine counts the number of occurrences of each journal title in the retrieved set, matches the journal titles with the positional information of the base map (stored in the file scopus.dbf; available from the same website), and then generates an input file for VOSviewer named "overlay.txt".

[2] The user can replace the classification based on VOSviewer with the one based on Blondel *et al.* (2008), by replacing the field named "cluster" with the values of the field "blondel" in the file "scopus.dbf," for example, in Excel. The file has to be saved back as a .dbf-file in the format of dBase III or dBase 4. (If this format fails in Excel, one can use OpenOffice or SPSS.) One can use any other classification system analogously by replacing the values in the field "cluster" in scopus.dbf. The program overlay.exe reads the values as provided in this field.

When using this file as input, VOSviewer highlights the journal names used in the download with the cluster colors in the map, whereas all other points are faded in grey. The size of each journal as a node is depicted proportionally to the $\log_4(n + 1)$ with $n$ as the number of occurrences.[3] (The "+1" is added to prevent single occurrences from being displayed, since log(1) = 0.) This same normalization was used by Leydesdorff *et al.* (2013) for WoS data. The maps can thus be compared directly despite the differences in the underlying data.

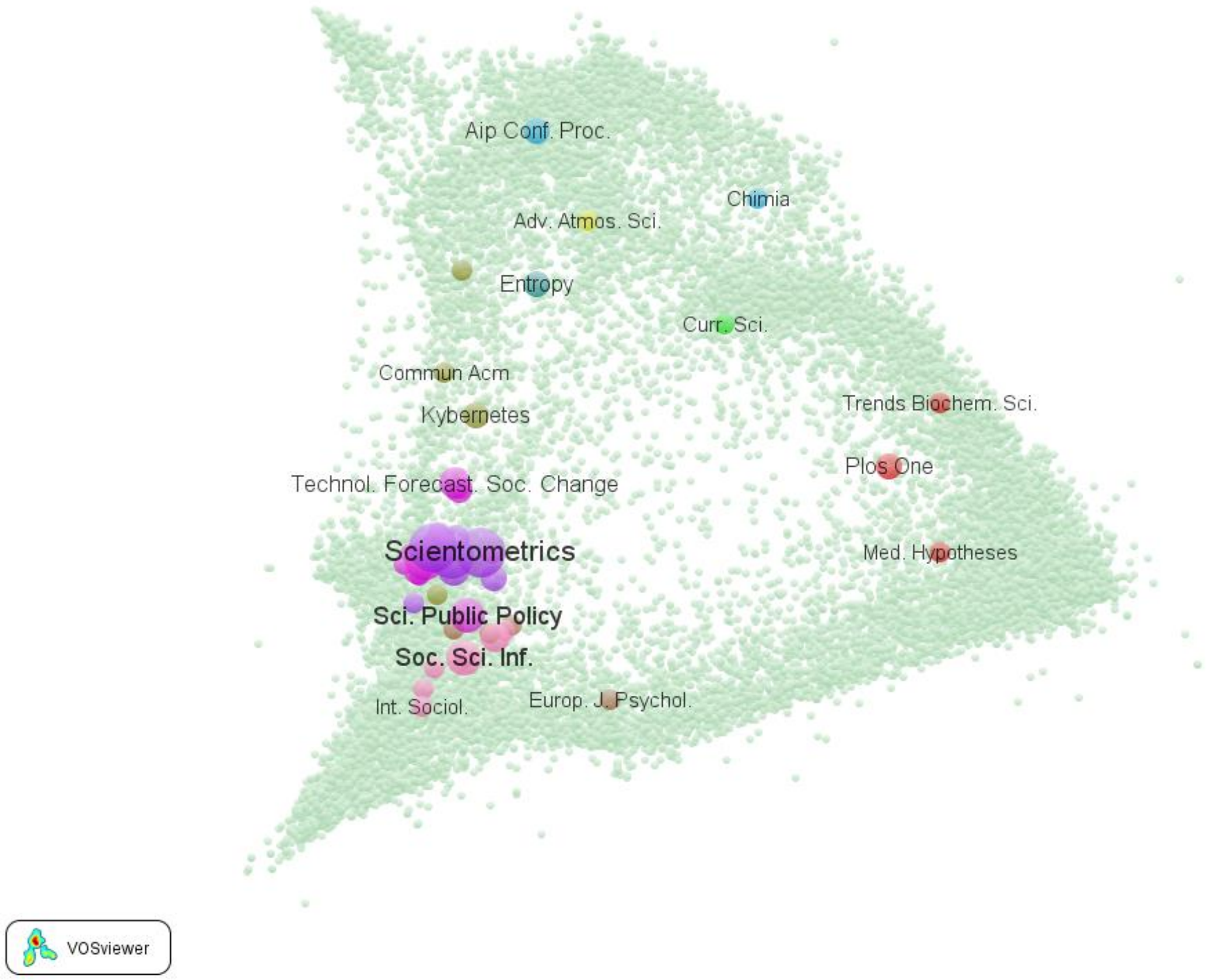

**Figure 5**: 244 documents retrieved from Scopus with the string 'AU-ID("Leydesdorff, Loet" 7003954276)' overlaid to the basemap; Rao-Stirling diversity: 0.068. Available at http://www.vosviewer.com/vosviewer.php?map=http://www.leydesdorff.net/scopus_ovl/ll.txt&label_size=1.35

[3] A file "overlay.dbf" is generated that contains all numerical information, such as the numbers of documents before the logarithm is taken.

Figure 5, for example, shows the overlay of 240 documents retrieved with the first author's author-identity in Scopus.[4] One can zoom in on the left bottom corner in VOSviewer and then make all labels visible. An instruction of how to generate an overlay is provided in an Appendix and also at http://www.leydesdorff.net/scopus_ovl/index.htm . The program overlay.exe writes the Rao-Stirling diversity value to a file "rao.txt" and to the screen before finishing. However, this file "rao.txt" is overwritten in each run and should thus be saved if one wishes to make comparisons. A file "overlay.dbf" is written at each run that contains the statistical information.

## 4. Applications

### *4.1 Digital and e-humanities*

One of the presumed advantages of using Scopus data is the inclusion of journals in the arts and humanities, whereas the *Arts & Humanities Citation Index* (A&HCI) of WoS has not been integrated into the system of Journal Citation Reports hitherto (see for a map of A&HCI: Leydesdorff, Hammarfelt, and Salah, 2011). In recent years, "e-humanities" has become a priority area in the advanced nations (Wyatt & Leydesdorff, in preparation). How is "e-humanities" represented differently in Scopus and WoS data?

We downloaded sets with the phrases "e-humanities," "ehumanities," "digital humanities," "computational humanities," and "humanities computing" from both databases on October 8, 2013, using only title words. (By using index terms or keywords, we would also evaluate the indexing in the two databases [Rafols & Leydesdorff, 2009].) The retrieval was 114 documents in

[4] The file was downloaded on October 8, 2013.

Scopus (in 71 journals) and 72 documents in WoS (in 38 journals), with Rao-Stirling diversity values of 0.124 and 0.069, respectively.

Figures 6a and 6b show that the larger retrieval in Scopus leads to a more informed overlay on the base map than that obtained by using WoS: 33 journal names are active in the left-hand Figure 6a against only 18 in Figure 6b on the right hand. (The right-hand map is based on the JCR 2011 data; for more details, see at http://www.leydesdorff.net/journals11 and Leydesdorff *et al*. [2013].) Obviously, the two solutions for the base lines are mirrored along the vertical axis. Although Rao-Stirling diversity cannot be compared precisely between the two mappings because of the different layouts, a factor two of difference is undoubtedly significant. In summary, both the retrieval and the interdisciplinarity of the sets were larger in Scopus than in WoS.

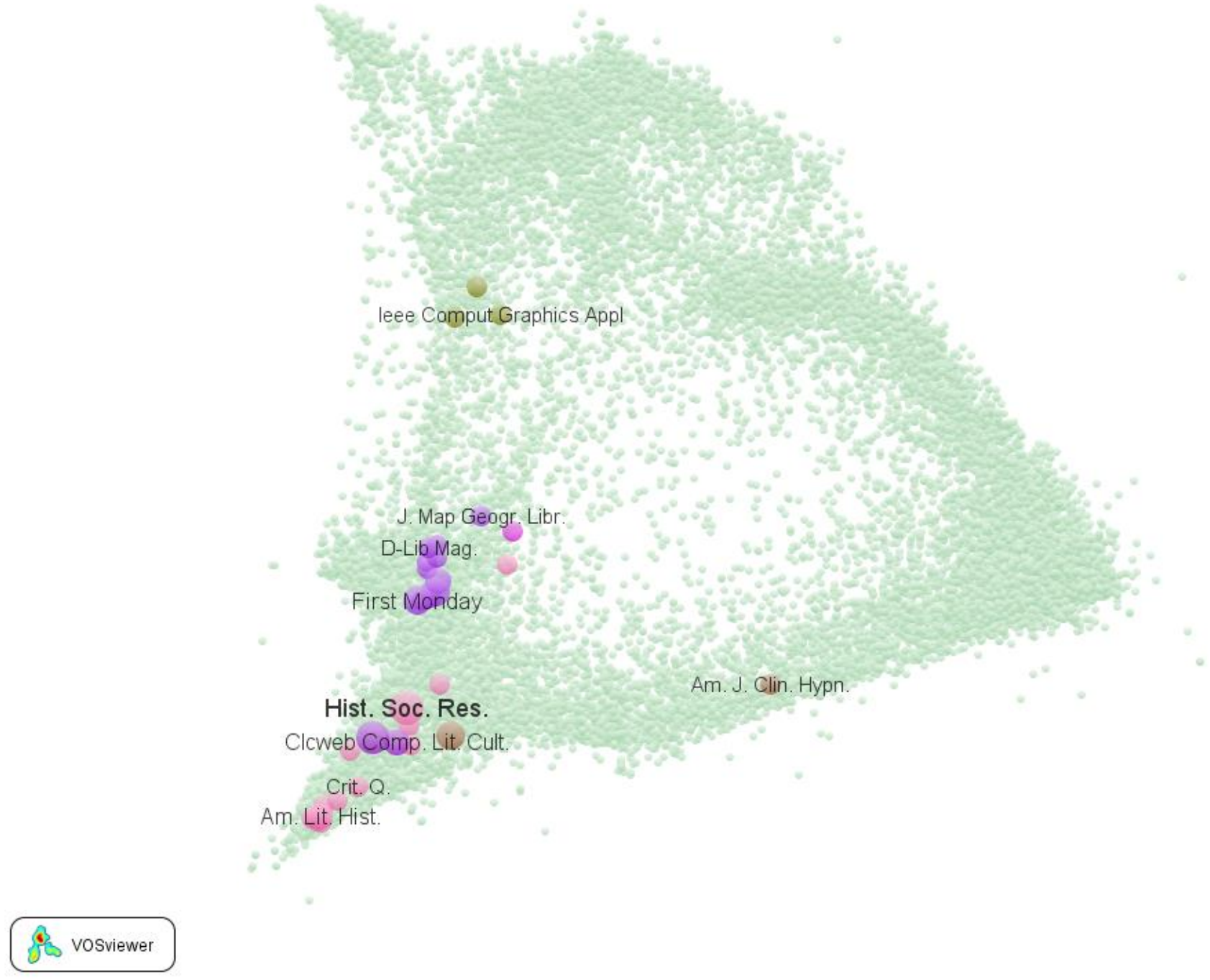

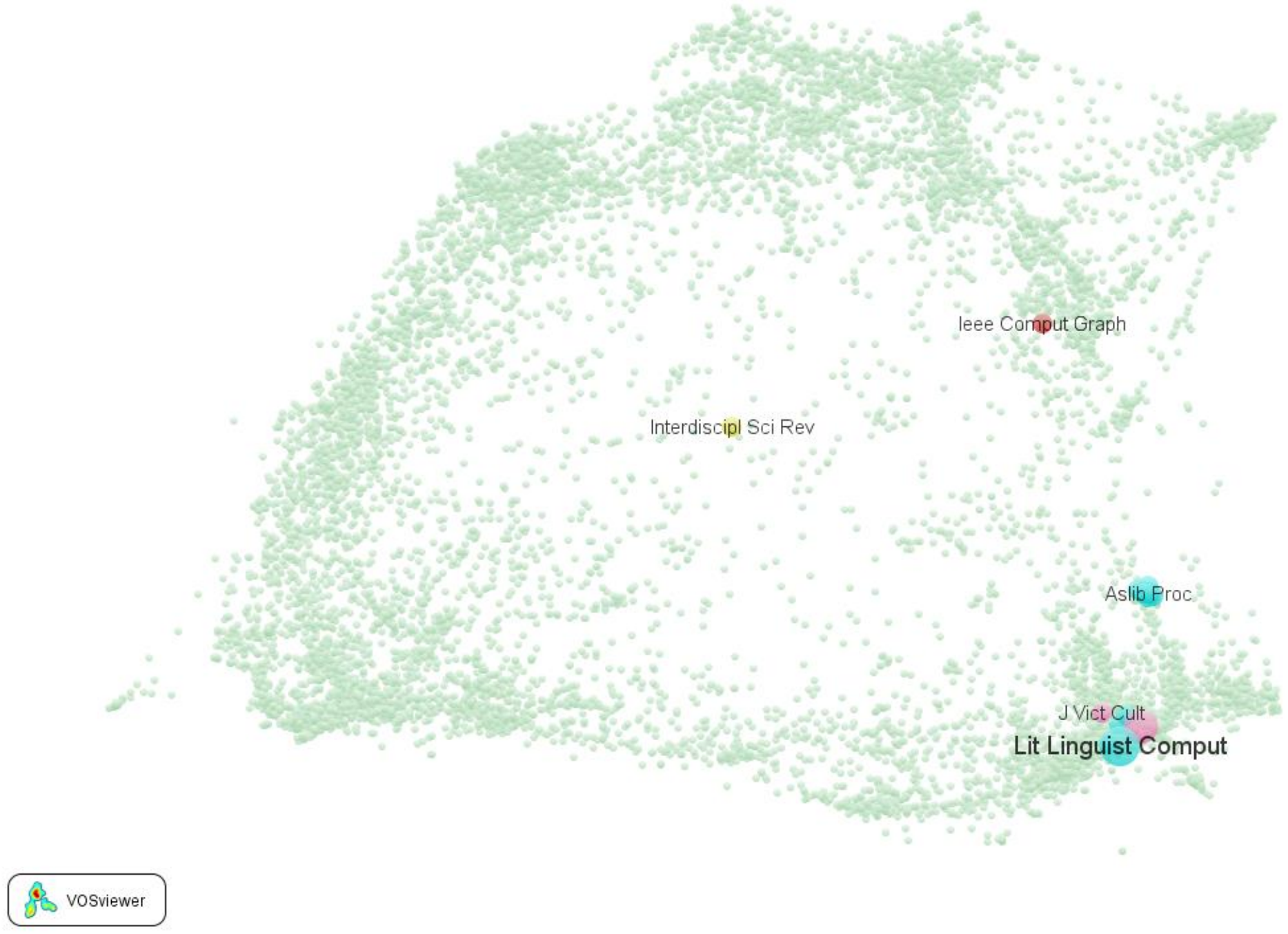

**Figure 6a** (left): Overlay of 114 documents with the search string 'TITLE("humanities computing") OR TITLE("computational humanities") OR TITLE("digital humanities") OR TITLE("ehumanities") OR TITLE("e-humanities")' in Scopus; Rao-Stirling diversity = 0.124; available at http://www.vosviewer.com/vosviewer.php?map=http://www.leydesdorff.net/scopus_ovl/ehum.txt&label_size=1.35

**Figure 6b** (right): Overlay of 72 documents with the search string 'ti="humanities computing" or ti="computational humanities" or ti="digital humanities" or ti="ehumanities" or ti="e-humanities'; All years. Databases=SCI-EXPANDED, SSCI, A&HCI. Rao-Stirling diversity = 0.069; available at http://www.vosviewer.com/vosviewer.php?map=http://www.leydesdorff.net/scopus_ovl/wos.txt&label_size=1.35

The two figures (6a and 6b) are very different in terms of the journals which are labeled. This seems remarkably inconsistent, but is caused by incidental differences in the retrieval rates at the level of individual journals between the two databases. In Scopus, for example, we retrieved ten documents from *Historical Social Research* while this number was only seven in WoS.[1] In WoS, however, we retrieved 16 documents from *Literary and Linguistics Computing*, whereas this number was 14 in Scopus. The differences may be caused by a number of factors such as the different years for inclusion of each journal in the database; differences in the definitions of publication and tape years, etc.

The two journals are positioned in both maps with approximately the same coordinates; however, the one label happens to be foregrounded in the map to the left and the other in the one to the right. If one zooms in to the maps, the two journals can both be made visible in both maps. Figure 7a shows this level of detail for the Scopus map. Having this almost similar position in the global maps of Figures 6a and 6b—so that the (overlapping) labels compete for showing in VOSviewer—does not mean that the two journals are similar in terms of their intellectual focus in their local citation environments.

Figure 7b shows the local map for *Literary and Linguistic Computing* which is very different from the local maps of *Historical Social Research*. The two journals belong to different intellectual specialties, but as journals in the humanities they are similarly positioned in the disciplinary structures of a global environment including the natural and life sciences. The two journals do not even appear in each other's direct citation environment when using local maps, yet they share an intellectual space in the humanities at the level of the global division of labour among the disciplines.

[1] In WoS, this journal is entitled *Historical Social Research – Historische Sozialforschung*.

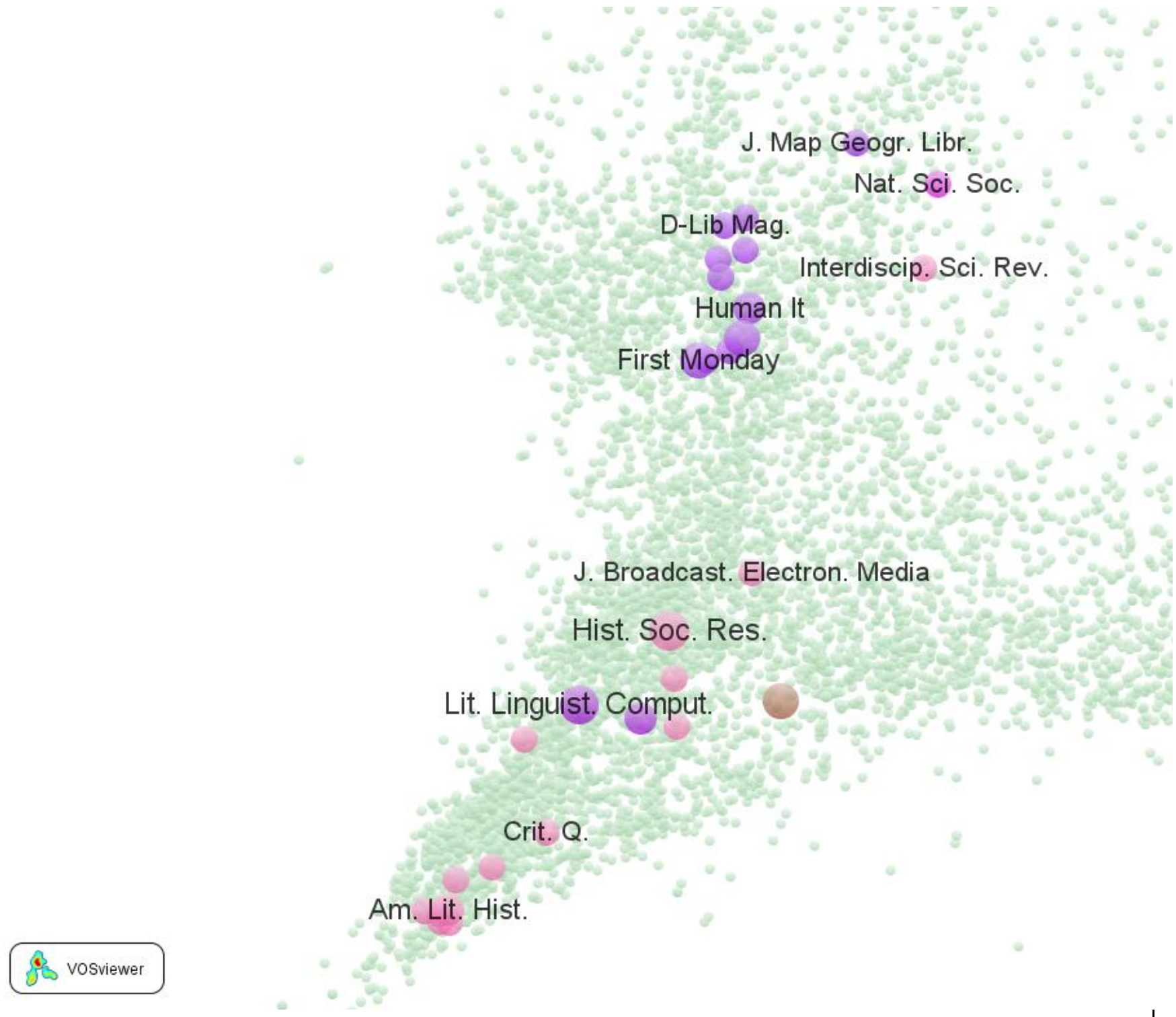

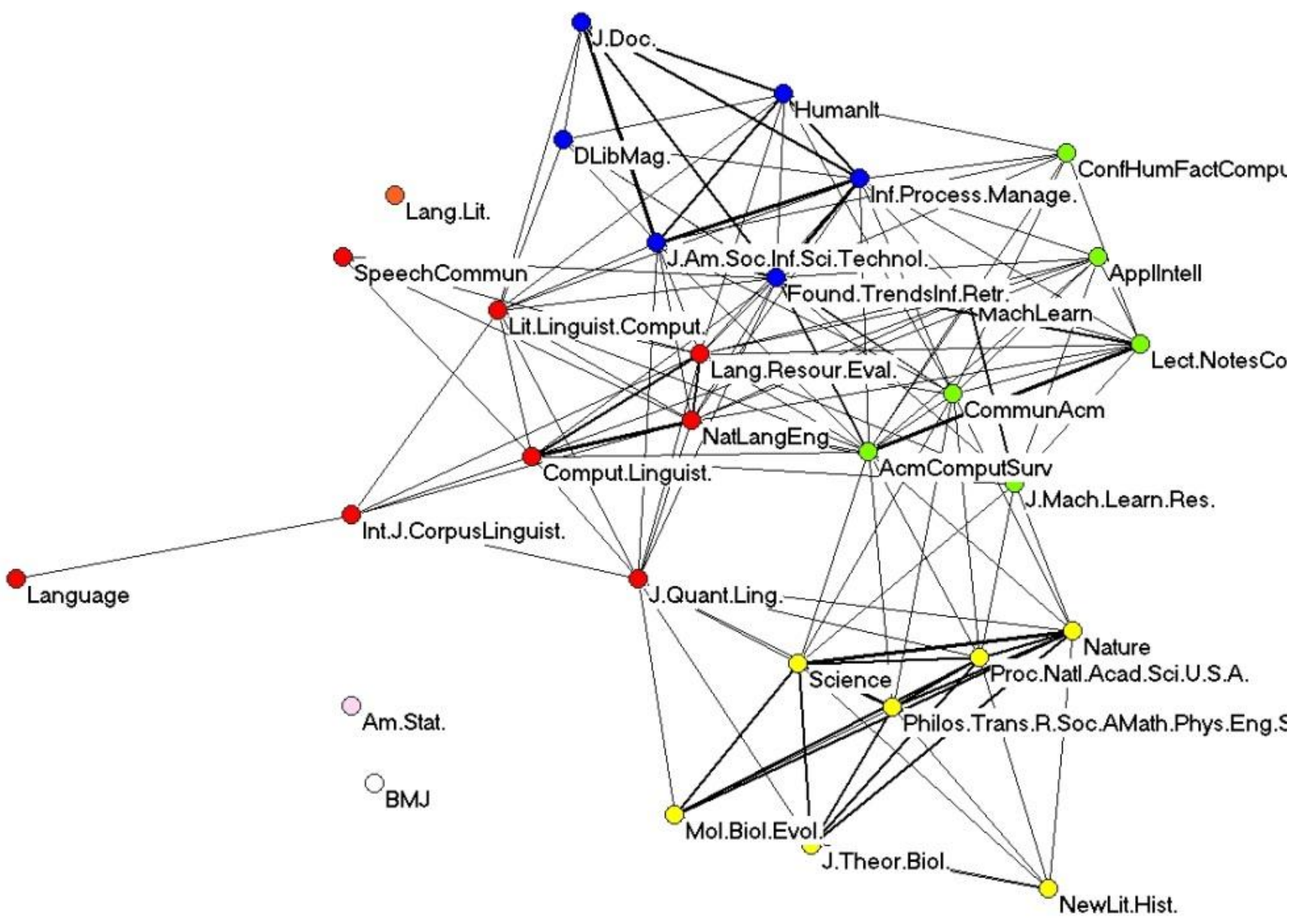

| | |
|---|---|
| **Figure 7a:** Enlargement of the relevant part of Figure 6a. | **Figure 7b:** local map of 31 journals contributing more than 0.5% to the total “citing” in *Literary & Linguistic Computing* in Scopus 1996-2012; $Q$ = 0.4555. |

### *4.2 Comparisons of portfolios*

In an argument about the evaluation of interdisciplinarity units, Rafols *et al.* (2012) showed that the Science and Technology Policy Research Unit (SPRU) at the University of Sussex is far more diverse in its portfolio than the London Business School (LBS). The same sets of documents used for this evaluation were also used by Leydesdorff *et al.* (2013) to show the respective portfolios projected on the base map of WoS (Figures 3a and 3b in that study). We repeated this same analysis here using the Scopus data.

In both cases, and using the same publication years (2006-2010), we retrieved approximately twice as many documents from Scopus as from WoS. Table 3 summarizes these differences in terms of both the retrieval and Rao-Stirling diversity. Figures 8a and 8b show the two overlay maps that are comparable to Figures 3a and 3b (using WoS data) in the previous publication.

| | *Scopus* | *WoS* |
|---|---|---|
| *SPRU* | 0.149 (*n* = 268) | 0.218 (*n* = 148) |
| *LBS* | 0.091 (*n* = 715) | 0.092 (*n* = 343) |

**Table 3**: Rao-Stirling diversity and numbers of publications for SPRU and LBS publications (2006-2010); mapped in the citing dimension using Scopus.

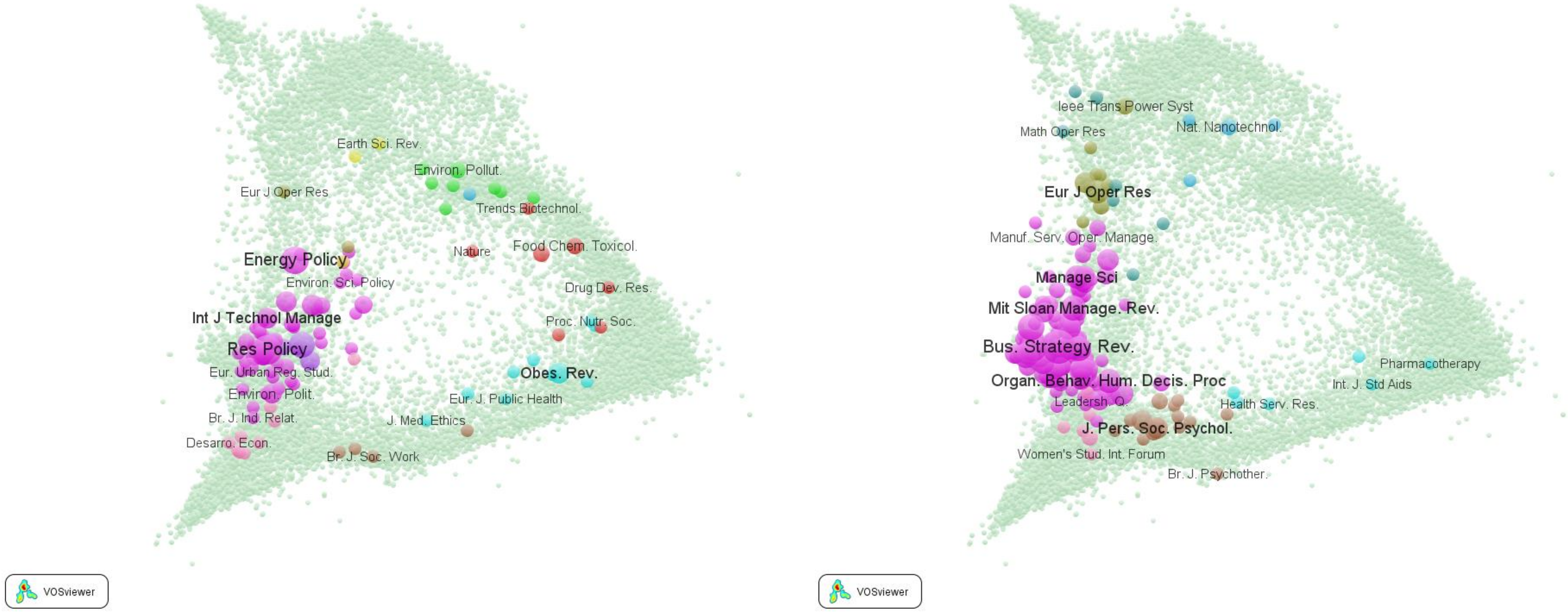

**Figure 8a and b**: Scopus-based overlay maps 2012 comparing journal publication portfolios from 2006 to 2010 between the Science and Technology Policy Research Unit SPRU at the University of Sussex (on the left; $N$ = 268; available at http://www.Vosviewer.com/Vosviewer.php?map=http://www.leydesdorff.net/scopus_ovl/spru.txt&label_size=1.35) and the London Business School (on the right; $N$ = 715; available at http://www.Vosviewer.com/Vosviewer.php?map=http://www.leydesdorff.net/scopus_ovl/lbs.txt&label_size=1.35).

As in the case of the e-humanities, the concentration is globally the same in terms of journals which belong to the same cluster (shown with the same colour), but the foregrounded labels are very different. Whereas in the case of SPRU, *Research Policy, Energy Policy,* and the *International Journal of Technology Management* are foregrounded, business strategy and management journals are most pronounced in the right-side figure for LBS.

In the case of SPRU, the spread across the map indicates the interdisciplinarity of the unit, whereas the output of LBS is concentrated in the disciplinary focus. Although the values of Rao-Stirling diversity are almost identical for LBS, this value was much higher for SPRU in WoS (Table 8). Thus, the inclusion of a larger set of journals in the case of Scopus leads to a more disciplinarily oriented representation of this research unit (SPRU). In other words, SPRU staff publishes also in disciplinarily core journals that are not included in WoS, but are included in Scopus.

In summary, the base map technique with interactive overlays and interdisciplinarity measurement can be most useful for the global analysis of portfolios and for comparative purposes. However, one may wish to follow up with local maps in order to analyze the intellectual organization at the specialty level because similar positions on the global map cannot show the fine structures of intellectual delineation at the specialty level (Zitt *et al.*, 2005).

## 6. Conclusions and discussion

When we explored the Scopus database for the purpose of journal mapping using the Scopus data for 2008, we did not find convincing evidence that this data would be richer than that of WoS; for example, when studying journals in the humanities or the social sciences (Leydesdorff, de Moya-Anegón, and Guerrero-Bote, 2010). The envisaged coverage of the humanities seemed at the time more a promise than a reality (Klavans & Boyack, 2009, at p. 464). Using the aggregated set 1996-2012, however, the difference from a JCR-based map is on the order of a magnitude. The inclusion of incidental citations to non-journal items such as dissertations may pollute the Scopus database, but this noise can be discarded by using a (relatively low) threshold.

The results are informative maps both locally and globally. We have shown this above with examples. However, there remains the limitation in Scopus of downloading maximally 2,000 records at each pass. (In WoS one can download in batches of 500, but with 100,000 as a maximum.) Because the data in WoS is more restricted than Scopus data in terms of journal delineations, the communities may be more discrete in WoS than Scopus data (see Figure 1). We also noted that a research unit (e.g., SPRU) can differently be appreciated using the larger context of Scopus when compared with WoS. Nevertheless, the two systems seem now in many respects roughly equivalent for the purpose of journal mapping. The maps for single years and the base maps do not contradict one another, but point to similar—torus-like—structures that were also found in other citation-based mapping efforts (Klavans & Boyack, 2009; cf. Bollen *et al.*, 2009).

In our description above we followed discursively the workflow from the premise on working toward a base map which would allow for overlays. Several choices were made among possible alternatives. Let us summarize the major decisions and parameter choices:

1. In addition to choosing between WoS and Scopus, one can choose between using an aggregate of years as in this study or to limit the analysis to the potentially thin layer of a single year;

2. The level of journals is a specific choice: large journals (e.g., PLoS One) may be interdisciplinary in themselves and therefore differently positioned in the network from year to year. Subject categories, however, may aggregate interdisciplinarity in the network at too high a level and introduce indexer effects (Rafols & Leydesdorff, 2009; cf. Rafols *et al.*, 2010);
3. We discussed the choice between "cited" as a representation of the common knowledge base versus "citing" as a representation of the currently shared citation environments;
4. We showed that without normalization, the larger-sized journals may overshadow the fine structures among smaller journals;
5. Among the many normalization possibilities (Waltman & Van Eck, 2013), we used the cosine values between vectors because the vector space provides us with a systems view that is consistent with the MDS-like procedure of VOSviewer (Leydesdorff, 2014);
6. The choice for VOSviewer was made because of the elegant solution of the problem of potential cluttering of large numbers of labels on screen;
7. We set informed default values to the overlay files that a user can generate, but indicate in the appendix how the user can modify these (e.g., the coloring and clustering of the journals; the scaling with the logarithm, etc.)

Some of the choices were pragmatically made, but the trade-offs were informed. For example, the use of (1 – cosine) as a distance measure in calculating the Rao-Stirling diversity would have been more consistent with the design, but potentially the interface would be overloaded with too large reference sets for calculating overlays efficiently. For this reason, we used the geometrical distances on the map as a measure. The difference between using these measures can be more systematically evaluated and made a subject for further research (cf. Leydesdorff & Rafols, 2011).

We have shown that journal mapping requires cosine-normalization of the citation data. Thus, one generates a vector space with properties different from those of the carrying networks of citation relations. In this vector space, all the journals are positioned. Without this normalization, the major journals dominate the map in terms of their "being-citedness," and the scattering is high on the citing side. Cosine-normalization solves these problems of concentration and scattering. The "citing" side then shows the current state of referencing a common knowledge base in the year of the download, whereas "cited" refers to the structure in and citation impact of the journal archive of the sciences.

Global maps and local maps show different and complementary things: two journals—as in the example above of *Literary and Linguistic Computing* and *Historical Social Research*—can be very different in terms of their integration at the specialty level, yet be equally positioned in the global structure as both belonging to the humanities. The projection of the multi-dimensional space on the two dimensions (*x* and *y*) of a map can be expected to generate stress. Unfortunately, stress values are not reported, but one should take this into account also when assessing the Rao-Stirling diversity measures; these are imprecise indicators, but one cannot easily specify the error in the measurement.

**Acknowledgement**

We are grateful to Lykle Voort of the Amsterdam computer center SARA for his support. Some of this work was carried out on the Dutch national e-infrastructure with the support of the SURF Foundation. We also thank two anonymous referees for their constructive comments.

## References

Blondel, V. D., Guillaume, J. L., Lambiotte, R., & Lefebvre, E. (2008). Fast unfolding of communities in large networks. *Journal of Statistical Mechanics: Theory and Experiment, 8*(10), 10008.

Bollen, J., Van de Sompel, H., Hagberg, A., Bettencourt, L., Chute, R., Rodriguez, M. A., & Balakireva, L. (2009). Clickstream data yields high-resolution maps of science. *Plos ONE, 4*(3), e4803.

Borgatti, S. P. (1997). Social Network Analysis Instructional Website, at http://www.analytictech.com/borgatti/mds.htm.

Boyack, K. W., Klavans, R., & Börner, K. (2005). Mapping the Backbone of Science. *Scientometrics, 64*(3), 351-374.

Boyack, K. W., Patek, M., Ungar, L. H., Yoon, P., & Klavans, R. (2014). Classification of individual articles from all of science by research level. *Journal of Informetrics, 8*(1), 1-12.

Fortunato, S. (2010). Community detection in graphs. *Physics Reports, 486*(3), 75-174.

González-Pereira, F.B., Guerrero-Bote, V.P., & Moya-Anegón, F. (2010). A further step forward in measuring journals' scientific prestige: The SJR2 indicator. *Journal of Informetrics, 4*(3), 379-391.

Guerrero-Bote, V.P., & Moya-Anegón, F. (2013). A further step forward in measuring journals' scientific prestige: The SJR2 indicator. *Journal of Informetrics, 6*(4), 674-688.

Kamada, T., & Kawai, S. (1989). An algorithm for drawing general undirected graphs. *Information Processing Letters, 31*(1), 7-15.

Klavans, R. & Boyack, K. W. (2009). Towards a consensus map of science. *Journal of the American Society for Information Science and Technology*, 60(3), 455-476.

Kruskal, J. B. (1964). Multidimensional scaling by optimizing goodness of fit to a nonmetric hypothesis. *Psychometrika, 29*(1), 1-27.

Lancho-Barrantes, B.S., Guerrero-Bote, V.P., & Moya-Anegón, F. (2010). What lies behind the averages and significance of citation indicators in different disciplines? *Journal of Information Science. 36*(3), 371-382.

Leydesdorff, L. (2014). Science Visualization and Discursive Knowledge. In B. Cronin & C. Sugimoto (Eds.), *Beyond Bibliometrics: Harnessing Multidimensional Indicators of Scholarly Impact* Cambridge MA: MIT Press.

Leydesdorff, L., & Cozzens, S. E. (1993). The Delineation of Specialties in Terms of Journals Using the Dynamic Journal Set of the Science Citation Index. *Scientometrics, 26,* 133-154.

Leydesdorff, L., De Moya-Anegón, F., & Guerrero-Bote, V. P. (2010). Journal Maps on the Basis of Scopus Data: A comparison with the Journal Citation Reports of the ISI. *Journal of the American Society for Information Science and Technology, 61*(2), 352-369.

Leydesdorff, L., Hammarfelt, B., & Salah, A. A. A. (2011). The structure of the Arts & Humanities Citation Index: A mapping on the basis of aggregated citations among 1,157 journals. *Journal of the American Society for Information Science and Technology, 62*(12), 2414-2426.

Leydesdorff, L., Kushnir, D., & Rafols, I. (2012; in press). Interactive Overlay Maps for US Patent (USPTO) Data Based on International Patent Classifications (IPC). *Scientometrics*, doi: 10.1007/s11192-012-0923-2.

Leydesdorff, L., & Rafols, I. (2011). Indicators of the interdisciplinarity of journals: Diversity, centrality, and citations. *Journal of Informetrics, 5*(1), 87-100.

Leydesdorff, L., & Rafols, I. (2012). Interactive Overlays: A New Method for Generating Global Journal Maps from Web-of-Science Data. *Journal of Informetrics, 6*(3), 318-332.

Leydesdorff, L., Rafols, I., & Chen, C. (2013). Interactive Overlays of Journals and the Measurement of Interdisciplinarity on the basis of Aggregated Journal-Journal Citations. *Journal of the American Society for Information Science and Technology, 64*(12), 2573-2586.

Leydesdorff, L., & Schank, T. (2008). Dynamic Animations of Journal Maps: Indicators of Structural Change and Interdisciplinary Developments. *Journal of the American Society for Information Science and Technology, 59*(11), 1810-1818.

Moya-Anegón, F., Chinchilla-Rodríguez, Z., Vargas-Quesada, B., Corera-Álvarez, E., Muñoz-Fernández, F. J., González-Molina, A., & Herrero-Solana, V. (2007). Coverage analysis of Scopus: A journal metric approach. *Scientometrics, 73*(1), 53–78.

Newman, M. E., & Girvan, M. (2004). Finding and evaluating community structure in networks. *Physical Review E, 69*(2), 026113.

Porter, A. L., Cohen, A. S., David Roessner, J., & Perreault, M. (2007). Measuring researcher interdisciplinarity. *Scientometrics, 72*(1), 117-147.

Rafols, I., & Leydesdorff, L. (2009). Content-based and Algorithmic Classifications of Journals: Perspectives on the Dynamics of Scientific Communication and Indexer Effects. *Journal of the American Society for Information Science and Technology, 60*(9), 1823-1835.

Rafols, I., Leydesdorff, L., O'Hare, A., Nightingale, P., & Stirling, A. (2012). How journal rankings can suppress interdisciplinary research: A comparison between innovation studies and business & management. *Research Policy, 41*(7), 1262-1282.

Rafols, I., & Meyer, M. (2010). Diversity and network coherence as indicators of interdisciplinarity: Case studies in bionanoscience. *Scientometrics, 82*(2), 263-287.

Rafols, I., Porter, A., & Leydesdorff, L. (2010). Science overlay maps: a new tool for research policy and library management. *Journal of the American Society for Information Science and Technology, 61*(9), 1871-1887.

Rao, C. R. (1982a). Diversity and dissimilarity coefficients: A unified approach. *Theoretical Population Biology, 21*(1), 24-43.

Rao, C. R. (1982b). Diversity: Its measurement, decomposition, apportionment and analysis. *Sankhy : The Indian Journal of Statistics, Series A, 44*(1), 1-22.

Schiffman, S. S., Reynolds, M. L., & Young, F. W. (1981). *Introduction to multidimensional scaling: theory, methods, and applications*. New York / London: Academic Press.

Stirling, A. (2007). A general framework for analysing diversity in science, technology and society. *Journal of the Royal Society Interface, 4*(15), 707-719.

Van Eck, N. J., & Waltman, L. (2010). Software survey: VOSviewer, a computer program for bibliometric mapping. *Scientometrics, 84*(2), 523-538.

Wagner, C. S., Roessner, J. D., Bobb, K., Klein, J. T., Boyack, K. W., Keyton, J., . . . Börner, K. (2011). Approaches to Understanding and Measuring Interdisciplinary Scientific Research (IDR): A Review of the Literature. *Journal of Informetrics, 5*(1), 14-26.

Waltman, L., & Eck, N. J. (2012). A new methodology for constructing a publication-level classification system of science. *Journal of the American Society for Information Science and Technology, 63*(12), 2378-2392.

Waltman, L., & Van Eck, N. J. (2013). A systematic empirical comparison of different approaches for normalizing citation impact indicators. *Journal of Informetrics, 7*(4), 833-849.

Waltman, L., & van Eck, N. J. (2013). A smart local moving algorithm for large-scale modularity-based community detection. *The European Physical Journal B, 86*(11), 1-14.

Waltman, L., van Eck, N. J., & Noyons, E. (2010). A unified approach to mapping and clustering of bibliometric networks. *Journal of Informetrics, 4*(4), 629-635.

Wyatt, S., & Leydesdorff, L. (in preparation). "E-Humanities" or "Digital Humanities:" Is that the question?

Zhou, Q., Rousseau, R., Yang, L., Yue, T., & Yang, G. (2012). A general framework for describing diversity within systems and similarity between systems with applications in informetrics. *Scientometrics, 93*(3), 787-812.

Zitt, M., Ramanana-Rahary, S., & Bassecoulard, E. (2005). Relativity of citation performance and excellence measures: From cross-field to cross-scale effects of field-normalisation. *Scientometrics, 63*(2), 373-401.

**Appendix**: How to create an overlay map using Scopus data?

One can generate an overlay for ***any set*** downloaded from Scopus as follows:

- Search in Scopus (advanced or basic); for example, using the search string 'TITLE("humanities computing") OR TITLE("computational humanities") OR TITLE("digital humanities") OR TITLE("ehumanities") OR TITLE("e-humanities")' provided 114 documents on October 8, 2013;
- Make a possible selection of records among the retrieved documents or tick "All";
- Click on "Export";
- Among the output formats, choose the "RIS format (Reference Manager, Procite, Endnote)" and "Specify fields to be exported";
- Only "Source titles" should be exported; untick all other fields;
- Click on "Export": the file "scopus.ris" can be saved; for example, for the 114 records mentioned at http://www.leydesdorff.net/scopus_ovl/scopus.ris;
- Save the file "scopus.ris" in the same folder as the routine **overlay.exe** (at http://www.leydesdorff.net/scopus_ovl/overlay.exe) and the file with the mapping information **scopus.dbf** (at http://www.leydesdorff.net/scopus_ovl/scopus.dbf; right-click for saving this file). One can paste different (e.g., sequential) output files of Scopus into a single file, but the routine expects an input file with the name "scopus.ris".
- One can now run overlay.exe in that same folder; preferably from the C-prompt; (using the C-prompt, one obtains error messages);
- The file "**overlay.txt**" (e.g., ehum.txt; at http://www.leydesdorff.net/scopus_ovl/ehum.txt ) is a map file that can be opened in **VOSviewer** (available at http://www.vosviewer.com );
- Rao-Stirling diversity is stored in the file "rao.txt" in the same folder, and shown on the screen; overlay.dbf contains the information such as the number of publications in each journal.

**Replacing the color and classification scheme**

The user can replace the classification based on VOSviewer with the one based on Blondel *et al.* (2008), by replacing the field named "cluster" with the values of the field "blondel" in the file "scopus.dbf," for example, in Excel. The file has to be saved back as a .dbf-file in the format of dBase III or dBase 4. (If this format fails in Excel, one can use OpenOffice or SPSS.) One can use any other classification system analogously by replacing the values in the field "cluster" in scopus.dbf. The program overlay.exe reads the values as provided in this field.

The output file "overlay.txt" can be opened by VOSviewer, and a map is then generated that can be adapted by the user using all the facilities present for its embellishment. The colors of the clusters can also be changed within VOSviewer. One can export in the various graphical formats (such as .png, .svg, or .eps). The map can also be made available for web-starting at the Internet. The user is advised to consult the manual of VOSviewer for further instruction. The program overlay.exe finally writes the Rao-Stirling diversity to a file "rao.txt" and to the screen before finishing. However, the file "rao.txt" is overwritten in each run and should thus be saved if one wishes to make comparisons. A file “overlay.dbf” is written at each run that contains the statistical information.